\documentclass{ws-procs9x6-cpt19}
\begin{document}

\newcommand{\refeq}[1]{(\ref{#1})}
\def\etal {{\it et al.}}

\newcommand{\bra}[1]{{\langle}#1\mid}
\newcommand{\ket}[1]{{\mid}#1\rangle}
\newcommand{\pbar}{$\bar{p}$}
\newcommand{\Hbar}{$\bar{\mathrm{H}}$}

\title{Status and Prospects for CPT Tests\\ with the ALPHA Experiment}

\author{T.\ Friesen}

\address{Department of Physics and Astronomy, University of Calgary,\\
Calgary, Alberta T2N1N4, Canada}

\author{On behalf of the ALPHA Collaboration}

\begin{abstract}
A primary goal of the ALPHA experiment at CERN 
is to perform precise tests of CPT symmetry. 
Here, 
we report on the significant progress made in recent years 
on antihydrogen spectroscopy and the outlook for the future. 

\end{abstract}

\bodymatter

\section{Introduction}

The ALPHA collaboration 
is focused on precision measurements with antihydrogen (\Hbar )
to test fundamental symmetries between matter and antimatter. 
The H--\Hbar\ system is natural for such tests 
because of its relative simplicity 
and its key role in the development of modern physics. 
Spectroscopy of H has a long and successful history: 
the 1S--2S transition 
and the ground-state hyperfine splitting (GSHFS)
have been measured at the levels of
$10^{-15}$ and $10^{-12}$, 
respectively.\cite{H1S2S,HHFS}
Since CPT symmetry 
implies the equality of the H and \Hbar\ spectra, 
reaching similar sensitivities in \Hbar\ 
represents an excellent CPT test 
in a purely atomic--anti-atomic system. 
Signals for CPT and Lorentz violation 
can be described using the Standard-Model Extension (SME).\cite{HbarSME}
The SME is a realistic effective field theory 
built from General Relativity, the Standard Model, 
and all possible Lorentz-violating operators.

\section{Antihydrogen trapping and detection}

The Antiproton Decelerator at CERN 
provides ALPHA with $3\times10^7$ antiprotons (\pbar ) every $\sim\!100\,$ s.
We capture roughly 90,000 of these in a Penning--Malmberg trap,
where we also load $3\times10^6$ $e^{+}$ from a Surko-type accumulator.\cite{Surko} The \pbar\ and $e^{+}$ are separately cooled, 
compressed, and then mixed to form roughly 50,000 \Hbar\ atoms. 
To confine \Hbar, 
ALPHA employs a magnetic-minimum neutral-atom trap 
formed by two short solenoids for axial confinement 
and an octupole winding for transverse confinement. 
In addition, 
there are three central short solenoids 
used to flatten the axial $\vec{B}$ field 
near the trap's center 
to aid with spectroscopy. 
Of the 50,000 atoms formed, 
an average of only 10--20 \Hbar\ 
have a low enough kinetic energy 
to be trapped. 
\Hbar\ atoms from consecutive mixing cycles can be accumulated, 
and hundreds of \Hbar\ atoms can be loaded into the trap 
in this manner.\cite{Stacking}

The annihilation of unconfined \Hbar\ 
is detected by a surrounding three-layer silicon annihilation detector. 
The \pbar\ annihilation on the Penning-trap electrodes 
will produce an average of three charged pions 
that register hits on each layer of silicon 
and allow the reconstruction of an annihilation vertex. 
The main background comes from cosmic rays 
that trigger the detector at a rate of $10 \pm 0.02\,$s$^{-1}$. 
Because the topology of the signal (annihilations) 
and background (cosmic rays) events are very different, 
they can be distinguished effectively 
by using machine-learning procedures.\cite{MVA}

\section{Antihydrogen Spectroscopy}

In the past several years, 
improved particle preparation techniques\cite{SDREVC,Stacking} 
have drastically increased \Hbar\ production and trapping rates at ALPHA,
opening up numerous avenues 
for performing experiments on \Hbar\  
including 1S--2S, hyperfine, and 1S--2P spectroscopy.
Below is a brief description of these efforts. 

ALPHA's primary spectroscopic goal has been 1S--2S spectroscopy because of the high precision achieved in H. 
The 1S--2S transition 
requires two simultaneous $243\,$nm photons and because we are interacting with small numbers of \Hbar\ 
(at most hundreds), relatively high power $243\,$nm radiation is needed. 
For these reasons, the ALPHA apparatus includes a cryogenic build-up cavity
that allows the input power of $160\,$mW to be built up 
to $\sim\!1\,$W of circulating power. 
After two-counter propagating photons 
excite the atom to the 2S state, 
absorption of a third photon ionizes the atom 
leading to loss and annihilation of the \pbar\ from the trap. 
In addition, 
atoms can couple from the 2S state to the 2P state 
and then be lost if they undergo a $e^+$ spin flip 
during decay back down to the 1S state. 
Both of these mechanisms lead to an annihilation signal
indicating that an excitation to 2S occurred. 
In 2018, 
ALPHA published a measurement
of the 1S--2S transition in \Hbar\ in a $1\,$T field 
to a precision of $5\times10^3\,$Hz out of $2.5\times10^{15}\,$Hz.\cite{Hbar1S2S} 
This is consistent with CPT invariance 
at a relative precision of $2\times10^{-12}$ 
(corresponding to an energy sensitivity of $2\times10^{-20}\,$GeV). 
This result has been used to put a constraint on CPT-violating SME coefficients, 
the first such constraint from \Hbar\ spectroscopy.\cite{ClockSME} 

Also of interest is the GSHFS in H, which is sensitive to different SME parameters 
than the 1S--2S transition 
and may potentially be even more sensitive to CPT-violating effects 
despite the lower relative precision of the measurement.\cite{HbarSME2}
The ground state of \Hbar\ in a strong $\vec{B}$ field 
is split into two pairs of states. 
One pair, 
the high-field seekers $\ket{a}$ and $\ket{b}$
with their $e^+$ spins aligned with $\vec{B}$, 
is not trapped by the magnetic-minimum trap. 
The other pair, 
the low-field seekers $\ket{c}$ and $\ket{d}$ 
with their $e^+$ spins anti-aligned with $\vec{B}$, 
is trapped. 
To measure the GSHFS, 
we excite the two positron spin-resonance transitions 
$\ket{c} \rightarrow \ket{b}$ and $\ket{d} \rightarrow \ket{a}$ 
and measure their frequencies. 
If both measurements are performed at the same $\vec{B}$ field, 
we can find the GSHFS through $f_{\mathrm{HFS}} = f_{da} - f_{cb}$. 
At a base field of $\sim\!1\,$T, 
these transitions occur at $\sim\!29\,$GHz. 
When an \Hbar\ undergoes such a transition, 
it is put into a high-field seeking state 
and will quickly annihilate on the surrounding apparatus walls. 
These annihilations are registered by the detector 
and used as the signal that a transition occurred. 

In 2017, 
ALPHA published the first measurement\cite{HbarHFS} of the GSHFS in \Hbar\ 
with a precision of four parts in $10^4$. 
Since that time, 
the rate at which trapped \Hbar\ can be produced and accumulated 
has increased considerably. 
Improvements have also been made to our ability to control, 
stabilize, 
and measure the trapping $\vec{B}$ fields. 
With more atoms and better field control 
this measurement can be improved significantly. 

A third transition that has been a major focus for ALPHA is the 1S--2P transition, 
which could be used to laser cool \Hbar\ for improved spectroscopy and gravity measurements. 
To excite this transition, 
narrow line-width (roughly $65\,$MHz) pulsed 
(about $12\,$ns duration) laser light at $121.6\,$nm 
is generated by third-harmonic generation of $365\,$nm light 
in a high-pressure Kr/Ar gas cell. 
Each pulse has an energy of $0.53$--$0.63\,$nJ in the trapping region, 
and the pulse repetition rate is $10\,$Hz. 
After being excited to 2P,
the atoms decay back to 1S within a few ns 
with a probability to undergo a $e^+$ spin flip 
and subsequently annihilate on the surrounding walls. 
In 2018, 
ALPHA published the results\cite{Hbar1S2P} of an experiment 
demonstrating the excitation of 1S--2P transitions. 
Based on a dataset consisting of 966 detected annihilations, 
ALPHA observed the 1S--2P transition in \Hbar\ 
and determined the transition frequency to a relative precision of $5\times10^{-8}$. 
With this demonstration, 
we are now in a position to attempt to laser cool \Hbar. 
Simulations predict 
that in the geometry of the current apparatus, 
laser cooling to roughly $20\,$mK is possible.\cite{HbarLC}

\section{Outlook}

After the major successes of recent years, 
ALPHA will continue to push \Hbar\ spectroscopy to new precisions, 
explore new transitions, 
and measure at different $\vec{B}$ fields. 
A further transition of interest is the \pbar\ spin flip transition between 
the $\ket{c}$ and $\ket{d}$ states. This transition exhibits a broad maximum near $0.65\,$T 
making it less sensitive to $\vec{B}$ fields, 
which is a major source of uncertainty. 
Two-photon 1S--2S spectroscopy in the near term 
can be improved by using larger waist size 
for the radiation in the optical cavity 
to reduce transit-time broadening. 
Also, laser cooling of the atoms down to $\sim\!20\,$mK 
would greatly narrow the measured linewidth. 
With the demonstration of our ability to excite the 1S--2P transition, 
laser cooling of \Hbar\ is within reach. 
Finally, 
ALPHA is also building a new apparatus 
to measure the gravitational free-fall of \Hbar. 
This new apparatus, 
known as ALPHAg, 
is a vertical magnetic trap 
that will initially allow us to determine if \Hbar\ falls up or down upon release 
and ultimately aims to measure the gravitational mass of \Hbar\ at the 1\% level.

\section*{Acknowledgments}
This work was supported by: the European Research Council through its Advanced Grant programme (JSH); CNPq, FAPERJ, RENAFAE (Brazil); NSERC, NRC/TRIUMF, EHPDS/EHDRS (Canada); FNU (NICE Centre), Carlsberg Foundation (Denmark); ISF (Israel); STFC, EPSRC, Royal Society, Leverhulme Trust (UK); DOE, NSF (USA); and VR (Sweden).

\end{document}